\documentclass[preprintnumbers,amsmath,amssymb,prc,showpacs]{revtex4} 
\usepackage[english]{babel}
\usepackage[dvips]{graphics,graphicx}
\usepackage{epsfig}
\usepackage{amssymb,amsmath,amsthm,amsfonts,amssymb}

\begin{document}

\title{The region interior to the event horizon of
 the Regular  Hayward   Black Hole.}

\author{Ivan Perez-Roman $^{}$}

\author{Nora Bret\'on$^{}$}

\affiliation{$^{1}$ Dpto de F\'isica, Centro de Investigaci\'on y de Estudios Avanzados \\
del I.P.N., Apdo. 14-740, Mexico City, Mex.}

\begin{abstract}
The Painlev\'e-Gullstrand coordinates allow us to explore the interior of  the regular Hayward black hole.  The behavior of an infalling particle in traversing the Hayward black hole is compared  with the one inside the Schwarzschild and Reissner-Nordstrom singular black holes. When approaching the origin the test particle trajectories present differences depending if the center is regular or singular. The velocities of the infalling test particle into the  modified Hayward black hole are analyzed as well. As compared with the normal Hayward, in the modified Hayward black hole the particle moves faster  and the surface gravity is smaller.
\end{abstract}


\maketitle

\section{Introduction}

The existence of a central singularity inside a black hole has been questioned as an incompleteness of the general relativity theory.  However  quantum cosmology points to a solution; for instance in loop quantum gravity (LQG)  has been resolved the big bang singularity replacing it by a quantum bounce and a recollapse \cite{Ashtekar2006}; quantum geometry modify the classical theory in such a way that, at least in some models like Friedman-Robertson-Walker, spacelike general relativity singularities are resolved \cite{Ashtekar2007}.

Moreover in \cite{Goswami2006} by studying the collapse of a scalar field, it is shown that  loop quantum gravity nonperturbative semiclassical modifications near the singularity, give rise to a strong outward flux of energy  that,  as a consequence,  prevents the formation of the singularity.

There are also string scenarios where matter originates from a decaying brane at the origin of time,  and it  can be interpreted as nonsingular initial conditions of the Universe \cite{Kawai2006}.

On the other hand, it is well known that at high densities of  matter, quantum effects become important, and the matter pressure may be able to counterbalance gravitational collapse. For instance neutron stars are supported by degeneracy pressure of electrons due to the Pauli exclusion principle.
It has been also proposed the existence of quark stars characterized by a small electro-weak core that can balance gravity \cite{DeChang2010}. Planck stars \cite{Rovelli2014} are also a promising proposal to solve the singularity issue as well as to give some insight  on the lost information paradox.

Therefore,  it seems reasonable to guess that most likely when matter reaches Plack density, that is the onset of quantum gravity effects,   there would be enough pressure as to prevent the formation of a singularity.  This situation has motivated the derivation and study of non-singular or regular black holes, starting with the seminal model  by Bardeen in 1968 \cite{Bardeen} of a  monopole characterized by a magnetic parameter whose inclusion allows to avoid the singularity at the origin.  Afterward regular black holes have been derived by including several kinds of matter, for instance, nonlinear electrodynamics, exotic fluids or  a central core sometimes modeled as a de Sitter region  \cite{Mbonye2005}, \cite{Spallucci2017}. 

 
For the reasons exposed above, it is interesting to explore the region interior to the event horizon of a regular black hole, particularly from the point of view of an infalling observer. The interior of regular black holes has  been explored, for instance regarding the thermodynamical properties of the matter inside the horizon \cite{SEPB2013}.  To traverse the horizon and to dig into the region beyond the horizon a special coordinate system is needed, one that is not  affected by the event horizon. One of these frames was introduced by Eddington and Finkelstein (EF)  to study the Schwarzschild spacetime at the event horizon; these coordinates are based on freely falling photons and are extensively used in discussions of gravitational collapse. In this paper  the  generalization of the EF coordinates is considered, the Painlev\'e-Gullstrand (PG) coordinates \cite{Poisson2001}  to explore the interior of a regular black hole.

Among the interesting regular black holes, Hayward (2006)  \cite{Hayward2006} proposed a static spherically symmetric black hole that near the origin behaves like a de Sitter spacetime, its curvature invariants being everywhere finite and satisfying the weak energy condition. Although this solution has been studied in several aspects, in this paper we analyze the situation inside  its event horizon. It turns out that the effect of the de Sitter center is  that of a repulsive force that brakes  the particle up to a complete rest at the center. The velocities and the time that takes to reach the horizon to the free-falling observer, as measured by a distant observer and in the proper time is compared.


The Hayward black hole possesses two horizons, a feature shared with the charged static spherically symmetric solution of the Einstein-Maxwell equations,  the Reissner-Nordstrom black hole. Comparison will be established between the  velocities measured by an infalling observer when traversing the regions limited by the two horizons in the Hayward and  the Reissner-Nordstrom (RN) spaces. In spite of being the  RN a charged one, while Hayward is uncharged, the Penrose diagram of both spacetimes is very similar, i.e. the causal structure of Hayward's metric is analogous to the one of the RN spacetime  \cite{Rovelli2015}. However several differences arise particularly as the particles approach the origin, signalizing the fact that Hayward's origin  is regular, unlike RN's center. 

Then  we shall consider  the modified Hayward's solution \cite{Rovelli2015}, that incorporates a time delay between
an observer at infinity and an observer in the regular center and  also takes into account some quantum effects. These effects are introduced through two parameters and the effect of varying these parameters in the velocities of the infalling particle is analyzed.  The  effective gravitational potential and the surface gravity are determined in terms of the black hole parameters. It turns out that there is a turning point at the origin and that the gradient of the potential, evaluated at the horizon,  is in proportion to the surface gravity.

The paper is organized as follows: in Section II we present the Hayward regular black hole and its horizons; in Section III  are determined the velocities of infalling test  particles in their own frame and in the one of a distant observer. In Section IV the comparison is established between the Hayward and the Reissner-Nordstrom black hole, as the test particle traverses the interior region. In Section V the modified Hayward black hole is introduced and the respective velocities of an ingoing particle are determined and compared with the normal Hayward black hole. Section VI deals with the effective potential and the  surface gravity felt by the test particle in traversing the interior region. Final remarks are given in the last section. In some phrases we may abbreviate black holes with BH.

\section{The regular Hayward black hole}
 
In \cite{Hayward2006} it was presented the formation and evaporation of a black hole, with an intermediate  quiescent phase of a static black hole so constructed to be regular  due to  an interior core. The metric is a static spherically symmetric one given by

\begin{eqnarray}
ds^2 &=& - F(r) dt^2+F(r)^{-1} dr^2+r^2 d\Omega^2, \nonumber\\
F(r)&=& 1-\frac{2M_{H}(r)}{r}, \quad M_{H}(r)= \frac{mr^3}{r^3+2mL^2}.
\label{Hayw_sol}
\end{eqnarray}

\noindent $m$ is a constant equal to the ADM mass at infinity,  $M(r \mapsto \infty)=m$.
The central core  is characterized with the parameter $L$, that has length dimensions,   whose effect is that of a repulsive force that  prevents  the singularity.  Near the origin the metric has  a de Sitter behavior, $F(r \mapsto 0) = 1- r^2/L^2+ O(r^5)$. Another consequence of including the repulsive core is that the strong energy condition might be violated, we shall return to this point in Section VI.  
If one makes the analog with the cosmological cases (for instance in loop quantum gravity  \cite{Ashtekar2007}), where the presence of a bounce prevents  the occurrence of the singularity, then such a repulsion should have its origin in quantum effects. For this reason  the parameter $L$ is associated  to the Planck length, that in terms of the universal constants is $L^2= G \hbar/c^3$ of the order of  $10^{-35}$m.  The constant $L$,  denoted in some works as $g^3=2mL^2$,  is also related with the remnant mass that is left after the process of evaporation reaches its final stage \cite{Ahmadi2018}; it turns out that the remnant mass is in proportion to the value of $g$.

The Hayward black hole possesses an outer and an inner horizon,  $r_{\pm}$.  The singularity can be avoided  because  $r_{-}$ is a Cauchy horizon, and then  lacking of global hyperbolicity, the singularity theorem by Penrose does not apply \cite{Rovelli2015}.
By setting $L=0$ the Schwarzschild solution is recovered.

\subsection{Horizons of the  Hayward Black Hole}

The radius of the inner and outer  horizons are solutions of  the Eq. $F(r)=0$, equivalently,

\begin{equation}
r^3-2mr^2+2mL^2=0.
\end{equation}

The third degree polynomial  has three real roots  if the discriminant $D$ is negative,

\begin{equation}
D= \frac{L^2m^2}{27} [27L^2-16m^2]<0.
\end{equation}

\noindent In this case
there is one negative and  two positive real roots; the latter being the ones that determine the positions of the external and internal horizons, given, respectively,  by

\begin{equation}
r_{\pm}= \frac{2m}{3}+ \frac{4m}{3} \cos \left[{ \frac{\pi}{3} \mp \frac{1}{3} \arccos \left({\frac{27L^2}{8m^2}-1} \right)} \right].
\label{horizons}
\end{equation}

\begin{figure}
\centering
\includegraphics[width=8cm,height=6cm]{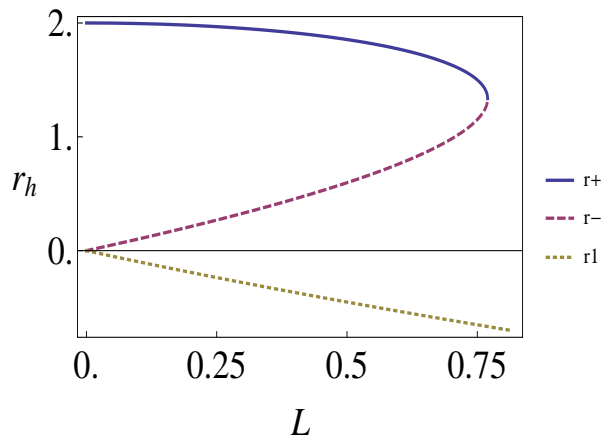}
\caption{\label{fig1} The three real roots of the third degree polynomial $F(r)=0$ that determine the horizons are displayed. The two positive roots define the outer (continuous curve) and inner (dashed) horizons. For the maximum allowed value of $L=4m/(3 \sqrt{3})$ the two horizons coalesce into one, at the turning point observed in the graphic. For $L=0$  one of the positive roots corresponds to  Schwarzschild horizon,  $r_{+}=2m$; the other one is  $r_{-}=0$.
In this plot $m=1$ and  the dotted line is the third (negative) real root $r_{1}$.}
\end{figure}

The three roots are shown as a function of $L$ in Fig. \ref{fig1}. The upper curve corresponds to the outer horizon  $r_{+}$, and the lower one to  the inner horizon $r_{-}$. The two intersections  between a vertical line at a given $L$  with the root-curves  mark  the positions of the inner and outer  horizons. Both horizons coalesce into one at the point where the lower and upper curves coincide, at the turning point of the parabola,  $r_{+} = r_{-} =4m/3$. This case corresponds to extreme BH in which $L=4m /(3 \sqrt{3}) \approx 0.77 m$ the maximum allowed value.  Note that  $D<0$  imposes a lower bound to the mass, $ m^2> \frac{27L^2}{16}$; or,  equivalently,  the sign of $D$  sets a restriction to the range of $L/m$ as $0 \le L/m \le 4/(3 \sqrt{3})$. 

\section{The motion of a free-falling observer}

Geodesics in the region exterior to the event horizon have been described  in \cite{Chiba2017}. We shall study the region interior to the event horizon considering a test particle impinging into the black hole from a very distant position, starting with zero velocity.
We shall derive the velocities of the free-falling object, both as measured by a  distant observer and  in the proper frame of the infalling  particle.

Because of the symmetries of the metric  (\ref{Hayw_sol})
the azimuthal component of the angular momentum and the energy of a test particle are conserved quantities.
A radial infalling trajectory is characterized by a zero angular momentum, then $ d \phi /d \tau =0$, where $\tau$ is an affine parameter. 
Another conserved quantity is $ds^2 /d \tau^2 = \delta$,  $\delta= -1, 0,1$ for timelike, null or spatial geodesics, respectively. Then  integrating  $dr / d \tau = \dot{r}$ from metric (\ref{Hayw_sol}),

\begin{equation}
\left({\frac{dr}{d \tau}} \right)= \mp \sqrt{1-F(r)},
\label{rdot}
\end{equation}
where  $ \dot{t}=dt/d \tau =- E g^{tt}=E/F$ has been substituted. The velocities of the ingoing observers will be taken  negative, considering that  $r$ is always decreasing.  In what follows the dependence of $F$ on $r$ may be omitted.

A distant observer measures the velocity of the infalling particle as $dr/dt$,

\begin{equation}
v=\frac{dr}{dt} = \left( \frac{dr}{d\tau} \right) \left( \frac{d \tau}{dt} \right) = - F \sqrt{ (1-F)},
\label{dist_vel}
\end{equation}

Derived from the previous velocities, the time intervals  measured in  the infalling  frame and  by the distant observer are, respectively,

\begin{equation}
d \tau = \frac{1}{ \sqrt{1-F}} dr,
\end{equation}
and
\begin{equation}
d t =  \frac{ dr}{F \sqrt{(1-F)}}.
\label{dt_distant}
\end{equation}

Since $0 \le F   \le 1$ and $0  \le (1- F)   \le 1$ the time interval measured  by a distant observer $dt$ is longer than the one measured in the particle's frame, $dt > d \tau$. In particular, as is well known, the time interval that takes a particle to reach the BH horizon is infinite  as measured by a distant observer ($F$ goes to zero in the denominator of Eq. (\ref{dt_distant})).

To determine the velocities in the infalling frame we use
the  Painlev\'e-Gullstrand (PG) coordinates. The PG coordinates were discovered independently by Painlev\'e (1921) and Gullstrand (1922)  for the Schwarzschild solution, and in a similar way than the Eddington-Finkelstein coordinates, or the Kruskal-Szekeres, they are well behaved at the event horizon, $r=2M$. Moreover, in the PG coordinates the free-falling particle is able to reach the origin $r=0$. Remarkable features of the PG coordinates are that the time coordinate is the proper time as measured by a free-falling observer starting from rest at infinity and moving radially inward; and that  the hypersurfaces with constant $t$ are all intrinsically flat.

To find  the expression of the  Hayward metric in  the Painlev\'e-Gullstrand coordinates, we define a new time coordinate as $ \tilde{t}=t-g(r)$ and,  using the condition that the spatial slices be flat,  we determine the function $g(r)$ as $g'(r)= \sqrt{(1-F)}/F$. Then the Hayward solution in Painlev\'e-Gullstrand coordinates has the form

\begin{equation}
ds^2=-F d{\tilde{t}}^2 -2 \sqrt{(1-F)} d \tilde{t} dr + dr^2 +r^2 d \Omega^2,
\label{Hayw_PG}
\end{equation}
 $\tilde{t}$ will be used to distinguish the PG-time coordinate from $t$,   the one measured by the distant observer. 
The metric can also be written  as

\begin{equation}
ds^2=-d\tilde{t} ^2  + \left( dr - \sqrt{(1-F)} d \tilde{t} \right)^2 +r^2 d \Omega^2.
\end{equation}

\begin{figure}
\centering
\includegraphics[width=10cm,height=6cm]{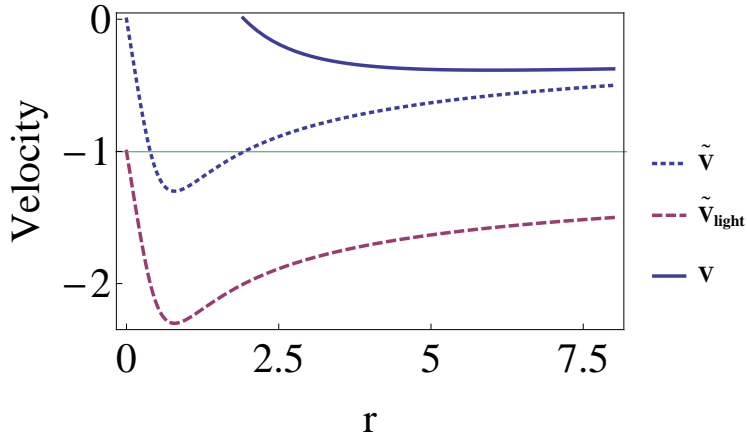}
\caption{\label{fig2} The velocities of a free-falling particle into Hayward  black hole, as seen by a distant observer (continuous curve) and by the one attached to the particle (dotted) are shown. The distant observer measures that the particle impinges the event horizon at  zero velocity and then loses contact.  In the frame of the infalling particle, it reaches the horizon with the speed of light ($\tilde{v}=-1$), then its velocity increases (in absolute value) up to a maximum (a minimum in the plot) and then starts to slowing down reaching the inner horizon with light speed; it continues slowing down approaching the center and arriving at $r=0$ with zero velocity. The lower (dashed) curve is for a massless particle (light) that penetrates the horizon, crosses both horizons with speed $\tilde{v}_{\rm light}=-2$ and reaches the center with light speed $\tilde{v}_{\rm light}=-1$. }
\end{figure}

Let us consider a radial infalling trajectory with $\dot{\phi}= d \phi /d \tau =0$,  $\dot{\theta}= d \theta /d \tau =0$, 
 $\delta=-1$ and  $\dot{\tilde{t}}= d \tilde{t}/d \tau =- E g^{tt}=E$,   integrating for $ (dr / d \tilde{t}) = \tilde{v}$, 

\begin{equation}
\tilde{v} = \left( {\frac{dr}{d \tilde{t}}} \right)= - \sqrt{ 1-F},
\label{PG_vel}
\end{equation}
Eqs. (\ref{rdot}) and (\ref{PG_vel})  do coincide for the ingoing test particle, what confirms that the time coordinate $\tilde{t}$ is the proper time.  In contrast,  the velocity measured by a distant observer,  Eq. (\ref{dist_vel})  differs from $\tilde{v}$ by a factor $F$, that makes that at the horizon the distant observer measures the velocity to be zero.
The velocities are illustrated in Fig. \ref{fig2}  both in the system attached to the particle and as measured by a distant observer.

In the frame attached to the infalling particle, it reaches  the outer horizon $r_{+}$ at the speed of light $\tilde{v}=-1$;  then,  between the outer and inner horizon, the velocity  increases (in absolute value) up to a maximum (minimum in Fig. \ref{fig2}).
The maximum of  $|\tilde{v} |$  is reached when $d \tilde{v}/dr=0$

\begin{equation}
\frac{d \tilde{v}}{dr}= \frac{F'}{2 \sqrt{1-F}}=0,
\end{equation}
where the prime means derivative with respect to $r$. 
From the previous equation we determine the radius where $F'=0$, that turns out to be $r_0=(4mL^2)^{1/3}$;  in Fig. \ref{fig2}  it corresponds to the minimum of the velocity curve. Beyond this point 
the velocity starts slowing down reaching the inner horizon $r_{-}$ at the speed of light again, and then  continuing slowing down such that  it arrives at $r=0$  with $\tilde{v}=0$. Note that if the particle were started its journey at infinity with a non-zero velocity, i. e. with some kinetic energy, the guess is that it would bounce at the origin, because of energy conservation. We shall return to this point in Section VI.

From Eq. (\ref{PG_vel}) for  the velocity, 

\begin{equation}
\tilde{v}= - \sqrt{\frac{2mr^2}{r^3+2mL^2}},
\end{equation}
it can be seen that  it decreases  (in absolute value) as $L$ increases.  Correspondingly the time lapses for reaching and crossing the black hole interior increase with $L$;  i.e. if $L_0 >L_1$ then the corresponding  intervals $\Delta t_{L_0} > \Delta t_{L_1} $.  

The behavior for a massless particle, light or photon,  qualitatively is the same as  described above,  with the difference that it never stops, but arrives to the origin with the velocity of light.  The velocity for an infalling massless particle in the PG coordinates is given by

\begin{equation}
\tilde{v}_{\rm light}= \left( \frac{dr}{d \tilde{t}} \right)_{\rm light}=  - 1 - \sqrt{ 1-F},
\label{PG_vel_ph}
\end{equation}

One can be persuaded  that  velocities greater than light are not unphysical 
making reference to the model proposed by Hamilton (2008) \cite{Hamilton2008}, that considers PG coordinates  as the ones attached to the flowing space, as if space were a river flowing through the interior of a black hole. 
Thinking of spacetime  as a river flowing in a Galilean fashion,  then there is  no contradiction in exceeding the speed of light.  The explanation of the observer reaching superluminal velocities is that it moves along with the space, that being Galilean is allowed to reach superluminal velocities. With respect to the river, objects  move according to the rules of special relativity, evolving by a series of Lorentz boosts and  never faster than light \cite{Hamilton2008}. Regarding superluminal velocities, although this fact challenges causality, recall that special relativity does not exclude faster-than-light signaling at the kinematical level and causality violation does not necessarily occur.
In \cite{Visser2001} are studied the superluminal effects that take place in the Casimir vacuum, and it is shown that such effects are constrained in such a manner as to not automatically lead to causality violations;  there is also introduced a definition of {\it stable causality}.

It is worth to note that the principle of BH complementarity for observers in the region exterior to the BH would require to perform one-sided coordinate transformation,  at the expense of the emergence of an associated shell of matter located on the horizon. To deal with this problem, that is beyond the scope of this paper, see \cite{Blau2016}.

\section{Comparison with Reissner-Nordstrom black hole}

The velocities of the infalling particles in Hayward's black hole  resemble some features of the ones in Reissner-Nordstrom's.
The Reissner-Nordstrom (RN) spacetime corresponds to the static spherically symmetric solution of the Einstein-Maxwell equations.
RN metric has the form (\ref{Hayw_sol}) with $M_{\rm RN}(r)= m-Q^2/(2r)$, where $Q$ is the electric charge.
The Schwarzschild solution  is recovered when the charge is zero,  $Q=0$.

\begin{figure}
\centering
\includegraphics[width=16cm,height=6cm]{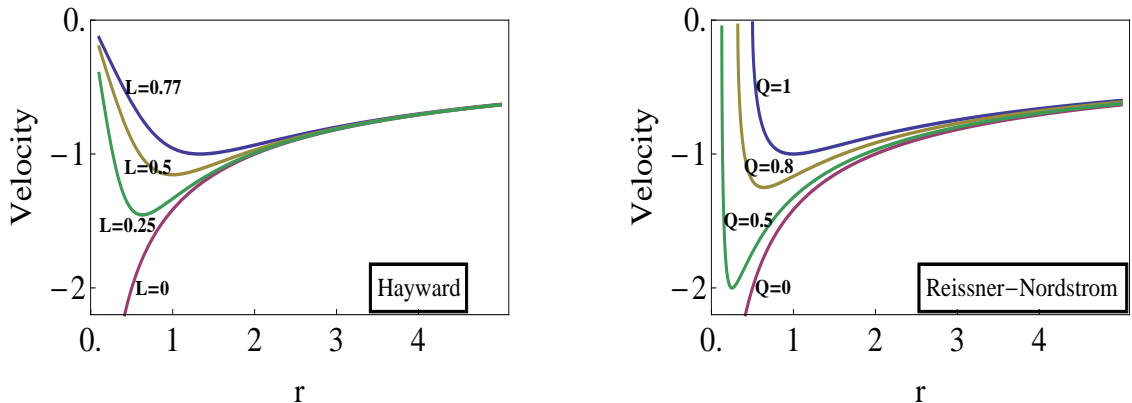}
\caption{\label{fig3} The velocities of a free-falling particle as seen by the observer attached to the particle are displayed for the Hayward  and the Reissner-Nordstrom (RN)  black holes,  varying in Hayward the parameter $L$ and the charge $Q$  in RN, with the values shown on the respective curves. In both black holes the particle reaches the event horizon with the speed of light ($\tilde{v}=-1$), and  afterward the speed increases (in absolute value) up to a maximum (minimum in the plot) and then starts to slowing down, reaching the inner horizon $r_{-}$ with speed of light.  As it approaches the center to Hayward's BH the particle slows down and  reaches the  center with zero velocity.  In contrast, for the RN case the velocity's particle  is zero before reaching the center, at $r_0=Q^2/2m < r_{-}$ and it never reaches $r=0$. The difference in trajectories as the particles approach the origin, is a manifestation of  a different center, in one case regular and  singular in the RN case. For $L=0$ or $Q=0$,  the motion in the Schwarzschild BH interior is recovered. In these plots $m=1$. }
\end{figure}

In the frame attached to the particle there is not much difference in the interior of the Hayward black hole and the one of the RN's, but the situation changes as the particle approaches the center.  In Figs. \ref{fig3} are shown the velocities of a test particle  falling into  a Hayward and a RN black hole,   varying $L$ in Hayward's case and the charge $Q$ in RN case, the corresponding values written on each curve.
In both cases the particle reaches the  horizon with the speed of light ($\tilde{v}=-1$, dashed horizontal line), and the speed increases  (in absolute value) until a maximum  (minimum in the plots) and then starts slowing down arriving at the inner horizon $r_{-}$ with speed of light. Then the particle continues slowing down its velocity.  From now on, the way of approaching  the  center is different. While in the Hayward's spacetime the particle arrives at the origin with zero velocity, in RN black hole  the velocity is zero at some point inside the inner horizon $r_0 < r_{-}$,  $r_0=Q^2/2m$, and beyond this point the particle's motion cannot be followed, at least in PG coordinates. The motion of an infalling test particle  in the RN black hole was analyzed in  \cite{Poisson2004}, concluding that inside the horizon  the gravitational potential  becomes repulsive due to the electric charge: Recall that $M_{\rm RN}=m-Q^2/2r$, then $M_{\rm RN}(r)$ becomes negative for $r$ small enough, producing then a repulsive gravitational potential.  In the RN case the metric can be continued   beyond the point where $\tilde{v}=0$, in another chart of coordinates, for instance in a Kruskal patch,   arising then a black hole tunneling, see details in \cite{Poisson2004}.

In summary, the motion of a test particle  is qualitatively  similar  in the interior of both black holes, the difference being near the origin:  in Hayward's the particle reaches the origin with zero velocity and, as we shall see in Section VI,  then it bounces, while in the RN case the velocity is zero before reaching $r=0$,  at $r_0=Q^2/2m < r_{-}$,  and beyond that point the particle's trajectory becomes unknown in PG coordinates.

\begin{figure}
\centering
\includegraphics[width=8cm,height=6cm]{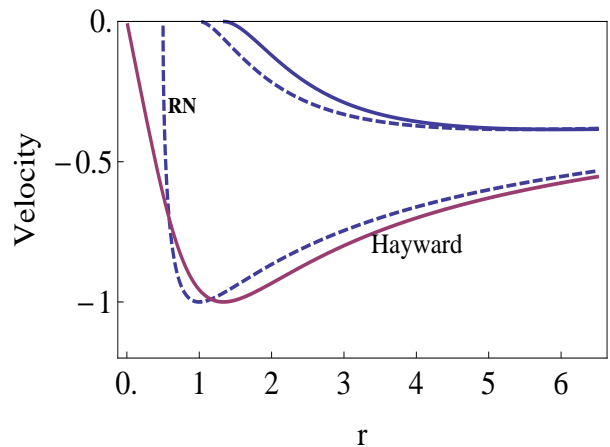}
\caption{\label{fig4} The velocities of a free-falling particle measured in the frame of a distant observer and in the one attached to the particle are shown, for the extreme RN (dashed) and extreme Hayward (continuous) black holes. The distant observer sees that the particle reduces to zero its velocity as it approaches the horizon. The RN horizon is located at $r_{-}=r_{+}=m$, while the Hayward horizon is at $r_{-}=r_{+}=4m/3$. In the frame of the infalling particle,  it reaches the horizon with the speed of light ($\tilde{v}=-1$) for both black holes and then starts  slowing down approaching the center, the difference being that  in the Hayward black hole, when the particle reaches the origin its  velocity is zero, while in the RN black hole the particle dissappears at $r_0=Q^2/2m=m/2$. }
\end{figure}

The extreme black hole, where the inner and outer horizons coalesce into one, in Hayward's occurs for $L= 4m/(3 \sqrt{3})$ and in RN for $Q=m$. In  Fig. \ref{fig4}  the velocities of a free-falling particle as seen by a distant observer and by the one attached to the particle are shown, comparing the extreme Hayward (continuous curves) and extreme RN (dashed curves) black holes. The distant observer sees that the particle reaches the event horizon with zero velocity, being the RN horizon located at $r_{-}=r_{+}=m$, while the Hayward horizon is at $r_{-}=r_{+}=4m/3$. In the frame of the infalling particle,   it reaches the horizon with the speed of light and then starts to slowing down approaching the center, the difference being that while in the Hayward black hole, the particle reaches the origin with zero velocity, in the RN BH the particle slows down to zero velocity before reaching $r=0$, precisely at $r_0=Q^2/2m=m/2$. According to some interpretations \cite{Poisson2004}, beyond $r_0$  the velocity changes to imaginary and the particle  tunnels  into another universe.   See \cite{Poisson1990} for a study on the interior of black holes and \cite{Hamilton2005} on the interior of charged black holes.

\section{The modified Hayward black hole}

It has been argued that the Hayward solution has two shortcomings, namely, it leaves aside the time delay  between an observer at infinity and an observer in the regular center and, secondly, it does not consider  the existence of a  quantum contribution at the center of the black hole.
In \cite{Rovelli2015}  the authors presented  a modification of the  Hayward metric that includes these two properties, incorporating a time delay as well as   1-loop quantum corrections to the Newton potential giving more support to the property that allows to avoid the singularity.  We  shall  call this metric the modified or improved  Hayward solution and  is given by

\begin{eqnarray}
ds^2 &=& - G(r)F(r) dt^2+F(r)^{-1} dr^2+r^2 d\Omega^2, \nonumber\\
G(r)&=& 1-\frac{ \alpha \beta m}{ \alpha r^3 + \beta m}, 
\label{Hayw_mod}
\end{eqnarray}
where $G(r)$ is  the correcting factor, chosen in such a way to preserve the physical sensible properties of $g_{tt}$ like asymptotic flatness. 
The metric component $g_{tt}=FG$ can be written as $1-2M(r)/r$,

\begin{equation}
FG=1- \frac{2M(r)}{r}=1- \frac{2}{r} \left[ { \frac{mr^3}{r^3+2mL^2}+ \frac{m \alpha \beta}{\alpha r^3+\beta m} \left(  \frac{r}{2} -
 \frac{r^3}{r^3+2mL^2} \right)}\right].
\end{equation}
The constant $\alpha$  incorporates a time delay between a clock at the center with respect to a clock at infinity; the values of $\alpha$ are in the interval $[0,1)$, recovering the normal Hayward solution when $\alpha=0$ .  The second parameter 
$\beta$ is for taking into account the 1-loop quantum corrections to the Newtonian potential, the suggested maximum value in \cite{Rovelli2015} is $\beta= 41/(10 \pi)$ . When $\beta=0$, then  $G(r)=1$ and the normal Hayward solution Eq. (\ref{Hayw_sol}) is recovered. The presence of $G$ does not modify the position of the horizons since the equation $G(r)=0$ possesses three real roots that are negative.

The velocities of a free-falling object  as measured by a  distant observer and  in the proper frame of the infalling  particle  in  the modified Hayward spacetime are given, respectively,  by


\begin{equation}
\frac{dr}{dt}= \pm F\sqrt{ G(1-GF)},
\label{dist_vel2}
\end{equation}

and

\begin{equation}
\frac{dr}{d \tau}= \pm \sqrt{\frac{1-GF}{G}},
\label{rdot2}
\end{equation}
where $+$ is for an outgoing particle and $-$ for an ingoing one.
Since we are focusing on the ingoing particles, we shall be taken the minus sign, that corresponds to an always decreasing coordinate``$r$".
Making $G=1$ Eqs. (\ref{rdot2}) and (\ref{dist_vel2})  reduce to  Eqs. (\ref{rdot}) and (\ref{dist_vel}), respectively.
Note that even when $L=0$  if $\alpha \ne 0$ and $\beta \ne 0$, the Schwarzschild case is not recovered,  corresponding then the metric to a Schwarzschild black hole with the retarding factor that accounts for the difference between the time at infinity and the one at the center of the black hole.

The modified Hayward metric in  the Painlev\'e-Gullstrand coordinates $(\tilde{t},r,\theta, \phi)$ has the form

\begin{equation}
ds^2=-FG d \tilde{t} ^2 -2 \sqrt{G(1-F)} d \tilde{t} dr + dr^2 +r^2 d \Omega^2,
\label{Hayw_modPG}
\end{equation}

The metric can be cast also as

\begin{equation}
ds^2=-G d \tilde{t} ^2  + \left( dr - \sqrt{G(1-F)} d \tilde{t} \right)^2 +r^2 d \Omega^2,
\end{equation}
in this way written, $F$ and $G$  can be related to the lapse and shift in the ADM formalism, being  $\sqrt{G}$ the lapse  and $ \sqrt{G(1-F)}$ the shift, \cite{Hamilton2005}, \cite{Hamilton2008}.

Determining  $dr / d \tau = \dot{r}$  from (\ref{Hayw_modPG}) gives

\begin{equation}
\frac{dr}{d \tau}=  \frac{1}{\sqrt{G}} \left\{ - \sqrt{1-F}  \pm \sqrt{1-G} \right\},
\label{PGrdot2}
\end{equation}
where the minus sign applies to the ingoing particle and the plus sign to the outgoing one.
And $dr/d \tilde{t}$ amounts to

\begin{equation}
\tilde{v}= \frac{dr}{d \tilde{t}}=  - \sqrt{ G(1-F)} \pm \sqrt{ G(1-G)}.
\label{PG_vel2}
\end{equation}
being the sign $\pm$ for outgoing and ingoing trajectories, respectively. 
The values of the metric functions at $r=0$ are $F(0)=1, \quad G(0)= 1- \alpha$, consequently, from the previous equation, the velocity at $r=0$ is

\begin{equation}
\tilde{v} (r=0) =  \pm \sqrt{ \alpha(1- \alpha)}.
\label{PG_vel2_r=0}
\end{equation}

In the modified Hayward spacetime, the time delay $G$  appears and the velocity $ ({dr}/{d \tau})$  Eq. (\ref{rdot2}) does not correspond to the one measured with the PG time $\tilde{t}$,   $ \tilde{v}=({dr}/{d \tilde{t}})$  Eq. (\ref{PG_vel2}). The property of spatial flatness at a slice of constant time is preserved,  however  there is a delay between the proper time and the PG time coordinate, given by $d \tau = G d \tilde{t}$.

\begin{figure}
\centering
\includegraphics[width=8cm,height=6cm]{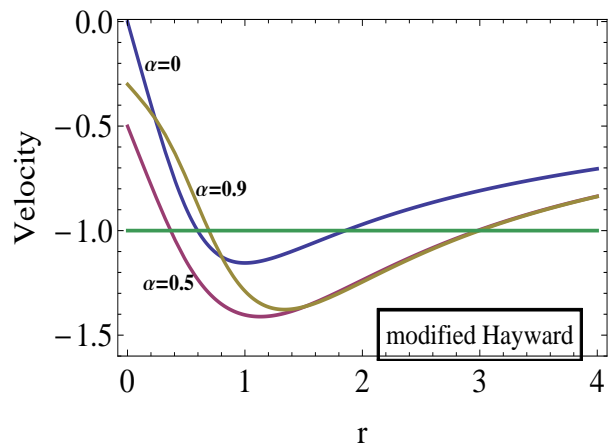}
\caption{\label{fig5} The velocities  $(dr/ d \tilde{t})$ of an ingoing particle into the modified Hayward black hole are shown for different values of the parameter $\alpha$, with $\beta=1.3$,  $m=1$  and $L=0.5$ fixed. Varying $\alpha$ changes the velocity, in particular, the test particle approaches the center with finite velocity,  $\tilde{v} (r=0)=\sqrt{ \alpha (1- \alpha)}$. }
\end{figure}

In Fig. \ref{fig5} the velocities of the free-falling particle as measured in PG coordinates, 
varying the parameter $\alpha$,  are plotted for fixed parameters
$\beta=1.3$,  $m=1$ and $L=0.5$.  The effect of increasing $\alpha$ increases the velocity;  particularly, the particle reaches the origin  with velocity
$ \tilde{v}(r=0)= \pm \sqrt{\alpha (1- \alpha)}$. The question then arises of what happens to the impinging particle, Does it dissappear or is there a bouncing and it returns along the same ingoing trajectory but in reverse? 

\begin{figure}
\centering
\includegraphics[width=14cm,height=6cm]{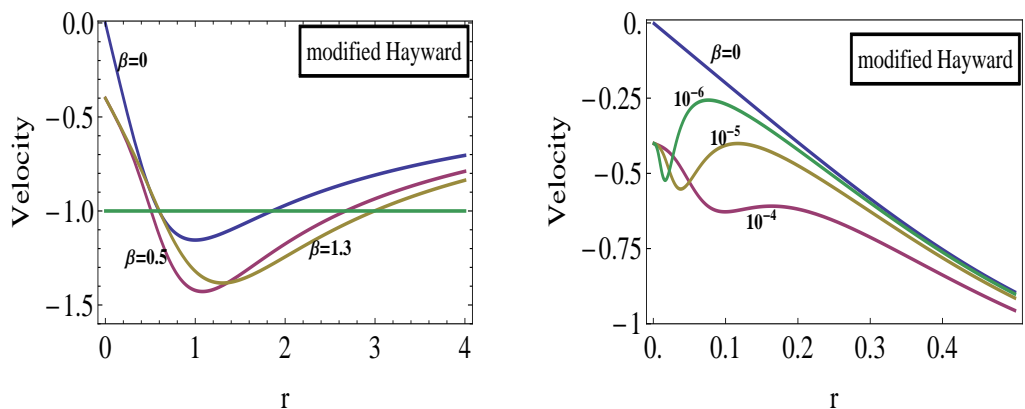}
\caption{\label{fig6}  Velocities  $(dr/ d \tilde{t})$ of a free-falling particle for the modified Hayward BH  are shown for different values of the parameter $\beta$. $\alpha=0.8$, $m=1$  and $L=0.5$. The normal Hayward BH is with $\beta=0$. The effect of introducing $\beta$ changes the maximum of the  velocity  (in absolute value)  as well as the velocity of reaching the horizon, but does not change the velocity of reaching the center. To the right is illustrated how occurs the process of decreasing  $\beta$ up to $\beta=0$; from lower to upper the values of $\beta$ are $\beta= 5 \times 10^{-4},  5 \times 10^{-5}, 5 \times 10^{-6}.$
  } 
\end{figure}

In Fig. \ref{fig6} the velocities are displayed for different values of the parameter $\beta$ with fixed  $\alpha=0.3$, $m=1$  and $L=0.5$.  The effect of 
modifying $\beta$  changes the velocity of reaching the outer horizon as well as its maximum (minimum in the plot), while the velocity of reaching the center depends only on the value of $\alpha$. In  Fig. \ref{fig6}  to the right is shown the way it approaches the center for very small 
$\beta$ and how  the transition occurs from $\beta \ne 0$ to $\beta=0$.  The normal Hayward BH corresponds to $\beta=0$.

Therefore the parameters $\alpha$ and $\beta$  modify the spacetime making it softer or less dense, or equivalently, diminishing the repulsive potential,  in such a way that the particle moves faster than in the interior of the normal Hayward black hole. 


The trajectories of massless particles or light rays are qualitatively the same as described above and are given, for the distant observer,  by

\begin{equation}
\left( {\frac{dr}{dt}} \right)_{\rm light}= -  F \sqrt{ G},
\label{light_PG_vel2}
\end{equation}
accordingly, the distant observer  measures that the photon reaches the event horizon with zero velocity. In the PG coordinates for the  modified Hayward we get for the velocity with which the photon travels, 

\begin{equation}
\left( {\frac{dr}{d \tilde{t}}} \right)_{\rm light}=- \sqrt{G(1-F)}  \pm  \sqrt{G},
\label{light_PGrdot}
\end{equation}
$\pm$ for the outgoing and ingoing, respectively.
The time delay makes that light velocity at the horizon be greater in magnitude (less than one in the plot) than in the normal Hayward BH.
\section{Effective Gravitational Potential and Surface Gravity}
 
On its journey through the black hole interior the particle follows a geodesic, but the changes in velocity may be interpreted as the effect of  a gravitational potential. We shall derive the  gravitational potential  and the surface gravity that  acts upon the test particle, considering the modified Hayward metric and then recovering the case of the normal Hayward  making $G=1$.

By means of the geodesic equation the gravitational potential $\Phi$ that the particle is acted upon can be calculated,

\begin{equation}
\frac{d^2 x^{\alpha}}{d \tau^2} + \Gamma_{\beta \delta}^{\alpha} \frac{d x^{\beta}}{d \tau} \frac{d x^{\delta}}{d \tau} =0.
\end{equation}
The nonvanishing  Christoffel symbols,  for $\theta$ and $\phi$ fixed,  are $\Gamma_{rr}^{r}= g^{rt} (g_{rt})_{,r}, \quad \Gamma_{tr}^{r}= g^{rt} (g_{tt})_{,r} /2, \quad \Gamma_{tt}^{r}= -g^{rr} (g_{tt})_{,r}.$
Substituting into the geodesic equation and using that $\dot{\tilde{t}}= d \tilde{t}/d \tau= 1/G$, as well as the Eq. (\ref{PGrdot2}) for $\dot{r}$  we obtain the equivalent to the gradient of a gravitational potential (a {\it fictitious  force}), $\nabla \Phi$, as measured in the system attached to the particle.  For the ingoing and outgoing  particles,  respectively $\mp$,  the geodesic equation amounts to

\begin{eqnarray}
0&=&\ddot{r} \mp \frac{ \sqrt{1-F} \sqrt{1-G}}{G^2}  \left\{  2(F'G+FG') -  G' \right\}+ \frac{ F'}{2G} \left\{  G-4+4F  \right\} + \nonumber\\
&& \frac{G'}{2G^2} \left\{ 2-5F+4F^2-G +FG\right\}= \ddot{r} +  \frac{d\Phi}{dr}. 
\end{eqnarray}

In the normal Hayward case,  that $G=1,G'=0$, the first and third terms vanish and  it reduces to

\begin{equation}
\ddot{r} -F' \left( \frac{3 }{2} -2F \right)=0.
\label{geod}
\end{equation}
From the previous equations  the gravitational potential is given by,

\begin{equation}
\frac{d\Phi}{dr}= -  \frac{mr(4mL^2-r^3)(2mL^2+r^3-8mr^2)}{(r^3+2m L^2)^3},
\label{grav_pot_grad}
\end{equation}
${d\Phi}/{dr}$ becomes zero at two points, where the factors in the numerator become null. First at $r_0=(4ML^2)^{1/3}$,  where the velocity reaches its maximum in absolute value,   and  at  $r=0$.
The third factor in the numerator does not become zero for the allowed values of $L/m \le 4/(3 \sqrt{3})$, i.e.   $(2mL^2+r^3-8mr^2)=0$  has no real roots.

Eq. (\ref{geod}) can also be written as

\begin{equation}
\frac{1}{2} \dot{r}^2 + \Phi = E,
\end{equation}
where $E$ is the energy of the test particle at infinity; from this equation we determine the turning points of the trajectory, $\dot{r}=0$, integrating  Eq. (\ref{grav_pot_grad}),

\begin{equation}
\Phi =  - \frac{mr^2(2mL^2+r^3-4mr^2)}{(r^3+2m L^2)^2}+  E,
\end{equation}
so there is  only one turning point at $r=0$. Therefore there is the possibility of a bounce  of the test particle at the origin.  There are no more turning points because the restriction on the values of $L/m$ does not allow  any real root for the equation $(2mL^2+r^3-4mr^2)=0$.  In the case of the modified Hayward it is not so easy to determine analytically the solution for the turning points, 
and the situation is more complicated since the particle reaches the origin with nonzero velocity.

At the horizon $r_{+}$, $F(r_+)=0$, and Eq. (\ref{geod})  reduces to

\begin{equation}
\ddot{r} - \frac{3}{2}{F'(r_+)}=0, 
\label{ddotr_hor}
\end{equation}
then we see that the gravitational potential that pulls the ingoing particle is of the same magnitude  than the one that drives the outgoing particle. 
This is related to the value of the surface gravity at the horizon as we shall see in the next subsection.

\subsection{Surface gravity}

The surface gravity is the acceleration of a static observer near the horizon, as measured by a static observer at infinity.
For the surface gravity $\kappa$, we adopt the definition  given in \cite{Visser2006},

\begin{equation}
\kappa(r)= \frac{1}{2 r} \left[{1-2M'(r)}\right]= \frac{1}{2r} \left\{ G(F+ r F') + rFG' \right\},
\label{surf_grav}
\end{equation}
that when evaluated at the horizon ($F(r_+)=0$) reduces to  $\kappa(r_+)=G(r_+) F'(r_{+})/2.$

\begin{figure}
\centering
\includegraphics[width=8cm,height=6cm]{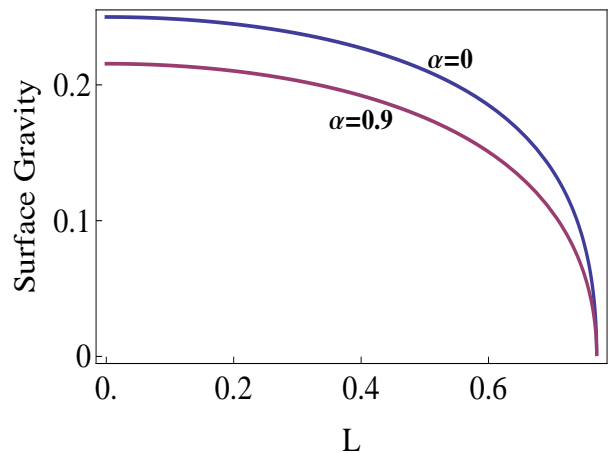}
\caption{\label{fig7}  It is displayed the surface gravity $\kappa$ for Hayward ($\alpha=0$) and the modified Hayward ($\alpha=0.9$) BHs, as a function of  $L$. The introduction of $L$ reduces the surface gravity; for $\alpha=0$  the largest surface gravity is the Schwarzschild's one, $\kappa_{Schwarz}=1/(4m)$; then diminishing as $L$ increases. Note that if $\alpha \ne 0$ even for $L=0$, $\kappa$ is smaller than Schwarzschild's. $\kappa=0$ for the extreme black hole that occurs for $L=4m/(3 \sqrt{3})$. $m=1$ in this plot. }
\end{figure}

From the expression for $\kappa$, Eq. (\ref{surf_grav}) and since $0 < G \le 1$ the following inequality holds at the horizon $r_{+}$,

\begin{equation}
\kappa_{H}= \frac{F'(r_+)}{2} \ge \frac{G(r_+) F'(r_{+})}{2}= \kappa_{\rm mH}.
\label{kapas}
\end{equation}
where subindex $H$ is for normal Hayward BH and $mH$ for modified Hayward BH.
Comparing  Eq. (\ref{kapas})  with the expressions for the gradient of the gravitational potential evaluated at the horizon,  $\nabla \Phi (r_+)$ from Eq. (\ref{ddotr_hor}),   we see a proportionality;  for the ingoing and the outgoing particle the surface gravity $\kappa_{H}= F'(r_+)/2$  is one third of  $ |\nabla \Phi (r_+)|$,  in agreement that that is the necessary force to hold a particle just outside the horizon.  
For the normal Hayward the surface gravity  amounts to

\begin{equation}
\kappa(r_+)= \frac{1}{2r_{+}}- \frac{3}{2} \frac{L^2}{r_{+}^3},
\end{equation}
then $\kappa(r_+)$ diminishes if $L$ increases. It is interesting how surface gravity changes as a function of $L$, 
evaluating  $\kappa$ at the horizon $r_{+}$ given by Eq. (\ref{horizons}), the behavior is shown in Fig. \ref{fig7}. For $L=0$  and $\alpha=0$ we recover the Schwarzschild result,  $\kappa=1/(4m)$, that is the departing point of the upper curve in Fig \ref{fig7}. The effect of increasing $\alpha$  is of decreasing the surface gravity;
$\kappa$ vanishes for  $L$ corresponding to the extreme black hole,  $L=4m/(3 \sqrt{3})$.  Recall that extreme black holes (one horizon) are characterized by their zero surface gravity (zero temperature). 

From Eq. (\ref{kapas}) we see that the thermodynamics of the Hayward and modified Hayward BHs will be different, since the associated temperature of the normal Hayward BH  is greater than the one for the modified BH.  The task is beyond the scope of this paper but  studies  related to the accretion of a fluid flow around the modified Hayward BH as well as its evaporation were presented in \cite{Debnath2015}.


Finally,  we derive,  for the normal Hayward solution ($G=1, G'=0$), the expression for the energy-momentum tensor by means of the (Bianchi-)Einstein tensor $G_{\mu \nu}$ and the Einstein equations.
This will give us an idea of the kind of matter that corresponds to such a core. Choosing an orthonormal tetrad comoving with the medium, for instance

\begin{equation}
\hat{t}^{a}=(1,-\tilde{v},0,0), \quad \hat{r}^{a}=(0,1,0,0),  \quad \hat{\theta}^{a},  \quad \hat{\phi}^{a}
\end{equation}

the orthonormal components of the Einstein tensor are

\begin{eqnarray}
G_{\hat{t} \hat{t}}&= & \frac{2M'(r)}{r^2}=  8 \pi T_{\hat{t} \hat{t}}= 8 \pi \rho, \nonumber\\
G_{\hat{r} \hat{r}}&=& - \frac{2M'(r)}{r^2}= 8 \pi T_{\hat{r} \hat{r}}= 8 \pi p_r, \nonumber\\
G_{\hat{\theta} \hat{\theta}}&= & - \frac{M''(r)}{r}=  G_{\hat{\phi} \hat{\phi}}= 8 \pi T_{\hat{\theta} \hat{\theta}}= 8 \pi p_{t}, 
\end{eqnarray}

Since $G_{\mu \nu}$ is diagonal and  $  G_{\hat{t} \hat{t}}=-G_{\hat{r} \hat{r}}$ and $ G_{\hat{\theta} \hat{\theta}}= G_{\hat{\phi} \hat{\phi}}$,
it is compatible with an energy-momentum tensor of an anisotropic fluid, $T_{\hat{a} \hat{b}}= $diag$[\rho, p_r, p_t,p_t]$.
It can be checked that such energy-momentum tensor satisfies the weak energy condition but violates the strong energy condition, meaning that
$\rho+p_r+2 p_t \ge 0 $ is not fulfilled for $r^3 \le mL^2$.  This violation is related to the de Sitter behavior of the metric near the origin, $F(r) = 1- r^2/L^2+ O(r^5)$.  Therefore the core matter is some kind of quantum non-isotropic fluid. Explicitly, the components of the energy-momentum tensor are,

\begin{eqnarray}
\rho &=& = - p_r= \frac{12m^2L^2}{8 \pi (r^3+2mL^2)^2}, \nonumber\\
p_{t} &=&  \frac{24m^2 L^2 (r^3-mL^2)}{8 \pi (r^3+2mL^2)^3}. 
\end{eqnarray}
The energy conditions for  the modified Hayward are also discussed in \cite{Rovelli2015}.

\section{Final  Remarks}

A thorough analysis and comparison between the Hayward  and Reissner-Nordstrom  black holes has been presented,  regarding the trajectories of infalling particles traversing the region interior to the event horizon. To that end the  Painlev\'e-Gullstrand coordinates have been used. The trajectories of the test particles are qualitatively  similar in both black holes, except when approaching the center. The effect of  increasing $L$ in Hayward black hole is similar to increasing $Q$ in RN, increasing the velocities.  Both black holes present an extreme case  (the two horizons merge into one), for  $Q=m$ in the RN BH, $r_{+}=r_{-}=m$, while for Hayward the extreme occurs if $L=4m/(3 \sqrt{3})$, being $r_{+}=r_{-}=4m/3$; the velocities for the extreme cases were illustrated as well.  The difference in trajectories occurs when approaching the origin, in Hayward's case 
the infalling observer has zero velocity when  reaches the origin,  in contrast with the situation in a singular black hole like Schwarzschild, that the test particle increases its velocity and it diverges as it approaches the center. In   Reissner-Nordstrom case the particle's velocity diminishes to zero at $r_0=Q^2/2m < r_{-}$ and then the particle cannot be tracked in PG coordinates.

Afterward it was analyzed the modified Hayward black hole, that takes into account the time delay and quantum effects through two parameters, $\alpha$ and $\beta$.  The effects of increasing these parameters increases the velocities of the infalling particles, equivalently diminishing the repulsive {\it force}  as if  the spacetime were less dense. For the external observer there is  no difference except in the surface gravity that decreases if $\alpha \ne 0$ and  $\beta \ne 0$.  

Comparing the surface gravity of the modified Hayward and the normal Hayward black holes, the introduction of $\alpha$ and $\beta$ diminishes the surface gravity. The changes in the velocities can be understand as the effect of a repulsive  gravitational potential whose gradient  turns out to be in proportion to the surface gravity at the  horizon.
For an external observer the signatures of the regularity of the  Hayward black hole are
the smaller surface gravity (see Fig. \ref{fig7})  as well as a smaller size of the horizon ( see Fig. \ref{fig1}) as compared with the  Schwarzschild black hole.
 
Regarding the fate of the test particle that reaches the origin $r=0$, according to our results, at $r=0$ there is a turning point, so the possibility exists of a bouncing.  However we could guess other possibilities like a tunneling through $r=0$ to a white hole \cite{Rovelli2014b}. To  assert  what really happens to the test particle require the analytic continuation through $r=0$ in other  coordinate system, but that study is beyond the scope of the present paper.

\begin{acknowledgments}
I. P-R acknowledges support by CONACYT through a Ms. C. fellowship. N. B. acknowledges partial  support by CONACYT Grant 284489.
\end{acknowledgments}



\begin{thebibliography}{
}

\bibitem{Ashtekar2006}
A. Ashtekar,  T. Pawlowski, P. Singh:
{\sl Quantum Nature of the Big Bang.},
{\em Phys. Rev. Lett.} {\bf 96} (2006) 141301. ArXiv: gr-qc/ 0602086

\bibitem{Ashtekar2007}
A. Ashtekar,  T.  Pawlowski, P. Singh,  K.  Vandersloot:
{\sl Loop quantum cosmology of $k=1$ FRW models.}
{\em Phys. Rev. D} {\bf 75} (2007) 024035. ArXiv: gr-qc/ 0612104

\bibitem{Goswami2006}
R. Goswami, P. S. Joshi, P. Singh:
{\sl Quantum Evaporation of a Naked Singularity.}
{\em Phys. Rev. Lett.} {\bf 96} (2006) 031302. ArXiv: gr-qc/0506129

\bibitem{Kawai2006}
S.  Kawai, E. Keski-Vakkuri, R.  G. Leigh, S. Nowling:
{\sl Brane Decay and an Initial Spacelike Singularity.}
{\em Phys. Rev. Lett.} {\bf 96} (2006) 031301. ArXiv: hep-th/0507163

\bibitem{DeChang2010}
De-Chang Dai,  A.  Lue, G. Starkmanb, D. Stojkovica:
{\sl Electroweak stars: how nature may capitalize on
the standard model's ultimate fuel.}, JCAP12(2010)004.
 ArXiv:0912.0520

\bibitem{Rovelli2014}
C. Rovelli,  F.  Vidotto: {\sl Planck Stars.},
Int. J.  Mod. Phys. D, {\bf 23}, (2014) 1442026 (11 pages).
 ArXiv: 1401.6562

\bibitem{Bardeen}
J. Bardeen. In Proceedings of the 5th International Conference
on Gravitation and the Theory of Relativity. Tbilisi, Georgia.
9–13 September 1968. Tbilisi University Press, Tbilisi. 1968.

\bibitem{Mbonye2005}
M. R. Mbonye, D.  Kazanas: {\sl Nonsingular black hole model as a possible product of gravitational collapse. }
{\em Phys. Rev. D} {\bf 72} (2005) 024016. ArXiv: gr-qc/ 0506111

\bibitem{Spallucci2017}
E. Spallucci,  A. Smailagic: {\sl  Regular black holes from semi-classical down to Planckian size.} 
{\em Int. J. Mod. Phys. D} {\bf  26} (2017) 1730013. ArXiv: 1701.04592

\bibitem{SEPB2013} D. P\'erez, G. E. Romero, C. A. Correa, S. E. Perez Bergliaffa: {\sl Analysis of a regular black hole interior.}
{\em  Int. J. Mod. Phys. Conf. Ser.} 2011.03:396-407. ArXiv:astro-ph/1111.0690.

\bibitem{Poisson2001}
K. Martel, E. Poisson: {\sl Regular coordinate systems for Schwarzschild and other
spherical spacetimes.}, 476 Am. J. Phys. {\bf 69},  (2001) 476-480. ArXiv: 0001069
 
\bibitem{Hayward2006} S. A. Hayward: {\sl Formation and Evaporation of Nonsingular Black Holes.}
{\em Phys. Rev. Lett.} {\bf 96} (2006) 031103. ArXiv: gr-qc/ 0506126

\bibitem{Rovelli2015} T.  De Lorenzo, C. Pacilio, C. Rovelli, S. Speziale, : {\sl On the Effective Metric of a Planck Star.}
{\em Gen. Relativ. Gravit. } {\bf 47} (2015) 41. ArXiv: gr-qc/ 1412.6015

\bibitem{Ahmadi2018}
S.  H.  Mehdipour, M.H. Ahmadi:
{\sl Black hole remnants in Hayward solutions and noncommutative effects}
Nuclear Physics B 926 (2018) 49–69. ArXiv: 1604.08584.

\bibitem{Chiba2017} Takeshi Chiba, Masashi Kimura, : {\sl A Note on Geodesics in the Hayward Metric.}
{\em Prog. Theor. Exp. Phys.} {\bf 2017}04E01.  ArXiv: gr-qc/ 1701.04910

\bibitem{Hamilton2008}  A. J. S. Hamilton and J. P. Lisle: {\sl  The river model of black holes.}
{\em Am J. Phys. } {\bf 76} (2008) 519-532. ArXiv: gr-qc/ 0411060

\bibitem{Visser2001}
S. Liberati,  S. Sonego, M. Visser: {\sl Faster-than-c Signals, Special Relativity, and Causality}, 
{\em Annals of Physics},  {\bf 298}, 167–185 (2002). ArXiv: gr-qc/ 0107091.

\bibitem{Blau2016}  M. Blau, M. O'Loughlin: {\sl  Horizon shells:Classical structure at the horizon of a black hole.}
{\em  Int. J.  Mod. Phys. D} {\bf 25} (2016) 1644010.  ArXiv: gr-qc/1604.01181

\bibitem{Poisson2004}
E. Poisson: A Relativist's Toolkit, Cambridge University Press (2004). Sect. 5.3.2.


\bibitem{Poisson1990} E. Poisson, W. Israel, {\sl Internal structure of black holes.}
{\em Phys. Rev. D} {\bf 41} 1796-1809 (1990).

\bibitem{Hamilton2005}  A. J. S. Hamilton and S. E. Pollack: {\sl Inside Charged  Black Holes. I. Baryons}
{\em Phys. Rev. D} {\bf 71} (2005) 084031. ArXiv: gr-qc/ 0411061

\bibitem{Visser2006} A. B. Nielsen, M. Visser : {\sl Production and decay of evolving horizons.}
{\em  Class. Quantum  Grav. } {\bf 23} (2006) 4637-4658. ArXiv: gr-qc/ 0510083

\bibitem{Debnath2015}
U. Debnath:  {\sl  Accretion and evaporation of modified Hayward black hole.}
{\em  Eur. Phys. J. C } {\bf 75} (2015) 129 ( 5 pages). ArXiv:1503.01645

\bibitem{Rovelli2014b}
H. M.  Haggard, C. Rovelli: {\sl Quantum-gravity effects outside the horizon spark black to white
hole tunneling.},
{\em Phys. Rev. D} {\bf 92} (2015) 104020. ArXiv: gr-qc/ 1407.0989

\end{thebibliography}
\end{document}